# Verifying Machine Learning Interpretability Requirements through Provenance

Lynn Vonderhaar
Dept. of Electrical Engineering
and Computer Science
Embry-Riddle Aeronautical
University
Daytona Beach, USA
vonderhl@my.erau.edu

Juan Couder
Dept. of Electrical Engineering
and Computer Science
Embry-Riddle Aeronautical
University
Daytona Beach, USA
ortizcoj@my.erau.edu

Daryela Cisneros
Dept. of Electrical Engineering
and Computer Science
Embry-Riddle Aeronautical
University
Daytona Beach, USA
cisnerd4@my.erau.edu

Omar Ochoa
Dept. of Electrical Engineering
and Computer Science
Embry-Riddle Aeronautical
University
Daytona Beach, USA
ochoao@erau.edu

*Abstract*—Machine Learning (ML) Engineering is a growing field that necessitates an increase in the rigor of ML development. It draws many ideas from Software Engineering and more specifically, from Requirements Engineering. Existing literature on ML Engineering defines quality models and Non-Functional Requirements (NFRs) specific to ML, in particular interpretability being one such NFR. However, a major challenge occurs in verifying ML NFRs, including interpretability. Although existing literature defines interpretability in terms of ML, it remains an immeasurable requirement, making it impossible to definitively confirm whether a model meets its interpretability requirement. This paper shows how ML provenance can be used to verify ML interpretability requirements. This work provides an approach for how ML engineers can save various types of model and data provenance to make the model's behavior transparent and interpretable. Saving this data forms the basis of quantifiable Functional Requirements (FRs) whose verification in turn verifies the interpretability NFR. Ultimately, this paper contributes a method to verify interpretability NFRs for ML models.

## I. INTRODUCTION

As Machine Learning (ML) models are continuously more heavily relied on, it is becoming critical that their development embodies the same level of rigor as that of traditional software systems. This is especially true as safety-critical domains, which tend to have stricter rules and regulations, explore the potential uses of ML [1]. Towards this goal, existing literature has drawn parallels between ML Engineering (MLE) and Requirements Engineering (RE), including defining ML-specific Non-Functional Requirements (NFRs) [2]. However, these definitions often lack a clear description of how the NFRs can be measured [2, 3]. Some literature notes that in addition to NFR measurement being difficult, decompositions from one NFR to another might also change for ML systems [3]. This work addresses both of these challenges for interpretability, which is defined in literature as an ML NFR, by decomposing it into provenance and using that provenance to implicitly verify it [2, 4].

Provenance is the origin of something [5, 6]. In the case of ML, it is the origin of model behavior. It includes various data throughout the model development process that contributes to the development of model behavior including datasets, preprocessing steps, and hyperparameters [4]. This contributes to behavioral transparency, which lends itself well to ML interpretability methods, e.g., anchors or decision rules. Therefore, this work proposes using provenance to decompose and verify model interpretability requirements by specifying which provenance data to save.

### A. Positioning within Existing Literature

This work contributes to RE for ML, so its contributions are requirements-based. It is important to note that although this work includes interpretability NFRs, it does not replace the interpretability methods found in literature. The purpose of this work is to define the provenance data necessary to use the standard interpretability methods found in literature. Defining this provenance provides a structured way to confirm that a system can be interpreted. This structure aligns with standard software verification techniques in RE. The contributions of this work are as follows:

1. Decomposing ML interpretability requirements into provenance requirements.
2. Verifying ML interpretability requirements using provenance Functional Requirements (FRs).

## II. BACKGROUND

Before describing this paper's approach, it is critical to first have clear definitions for RE, ML interpretability, and ML provenance.

### A. Requirements Engineering

RE is the systematic process of defining, documenting, and managing the requirements of a software or system. It ensures that the final product meets the needs of its stakeholders (customers, users, regulators, developers, etc.) and reduces the risk of building the "wrong" system [7]. RE is a crucial phase in the traditional Software Development Life Cycle (SDLC) because requirements act as the foundation for design, implementation, and testing. RE includes a series of activities which lead to requirement specifications, i.e., the formal, documented statement of what the system must do and the constraints under which it must operate. After these specifications are complete and finalized, they are used for Validation and Verification (V&V), ensuring the requirements are correct, complete, and reflect the stakeholders' needs [8].

RE for ML-based systems is still an emerging topic that requires additional application and adaptation of RE practices to the development of ML-based systems, with attention to data, model behavior, and uncertainty [9]. RE for ML is still the process of gathering, analyzing, documenting, validating, and managing requirements for ML-based systems, but it now handles not only functional needs, but also data quality, model performance, and fairness. At the end of this process, the requirements are still captured in requirement specifications, much like for traditional software [10]. However, the content of requirements specifications is different from those of traditional software as it must now capture data requirements, ML evaluation metrics, newly emerging NFRs, and system monitoring and retraining [11]. V&V is also still important in RE for ML-based systems, but literature notes that there are not yet methods for verifying some of the newly emerging NFRs [12, 2, 3]. Habibullah, et al. present a quality model for ML-based systems, but note that some ML-specific NFRs, e.g., interpretability, do not yet have measurements and are therefore difficult to verify [2].

*B. ML Interpretability*

ML model interpretability encompasses methods that allow model users to understand the underlying patterns within the data [2, 13]. This is distinct from explainability methods, which are means of understanding the model mechanics when making decisions. In other words, interpretability aims to provide insight into the data, while explainability provides insight into the model.

There are many existing interpretability methods and various surveys that organize them in unique ways [1, 14, 15, 16]. Some examples are:

- *Global versus local interpretability* where global interpretability methods show the model's generalized behavior, while local interpretability methods clarify reasons for a specific decision [14, 13],
- *Pre-hoc versus post-hoc interpretability* where pre-hoc interpretability design takes place before and during model training, while post-hoc interpretability methods are applied only after model training [1, 15],
- *Model-specific versus model-agnostic interpretability* where model-specific methods change based on the architecture of the black box model while model-agnostic methods do not [16], and
- *Explanation type* where interpretability methods are categorized based on their output, e.g., rule-based interpretability and saliency maps [17].

For the purposes of this work, interpretability methods are categorized by their underlying description mechanism so that a single example from each category will suffice to represent a provenance decomposition for each method in that category. The categorization of popular interpretability methods is shown in Table I. Table I is not meant to be a full survey of interpretability methods, but rather examples of popular methods for the purpose of verifying this work.

The interpretability methods in Table I are separated into three categories based on their underlying description mechanism: inherently interpretable models, surrogate models, and feature attribution. Inherently interpretable models have a simpler structure that allows developers and users to simulate their behavior [15]. Surrogate models are inherently interpretable models that train alongside more complex models and provide an approximate explanation for the complex model's behavior [14]. Feature attribution includes methods that directly describe the black box model's behavior with respect to input values [18]. While Table I is not a complete survey of interpretability methods, it provides some popular methods with which to verify this paper's approach.

*C. ML Provenance*

In its literal meaning, provenance means the origin of something [5, 6]. In ML, this refers to the origin, history, and transformations of data, models, and processes throughout the ML lifecycle [19]. In our previous work, we defined many types of provenance data in ML development, but this work focuses on two main categories: data and model [4]. The specific provenance in each of these categories is shown in Table II.

Data provenance is the record that describes the origins and processing of data. Data provenance helps with fairness, accountability, transparency, and explainability in AI systems [20]. Data provenance focuses on the lineage and history of data, answering questions about its source or origin, the methods of collection (such as human input, sensors, or web scraping), and the types of transformations applied, including cleaning, normalization, feature extraction, augmentation, or aggregation. It also captures who made modifications to the data, essentially tracing how the data changed and reached its current state through specific contexts or decisions [20]. Data provenance focuses on documenting the history of the data, from its origin to all the evolution and transformations until it reached its final state, so that it can be trusted, reused, and reproduced. Model provenance is very similar to data provenance, but it focuses on the model rather than the data. It captures the creation, origin, and evolution of the model by addressing questions related to its architecture, different versions, the changes it undergoes across those versions, the parameters and hyperparameters used, as well as the code, libraries, and environments involved [21]. Model provenance focuses on the history of how a model was built, trained, and modified, including its architecture, parameters, datasets, and environment, by keeping a record of all the changes the model itself has gone through.

TABLE I. CATEGORIZATION OF POPULAR INTERPRETABILITY METHODS. SYNTHESIZED INFORMATION FROM [1, 14, 15, 16, 13, 35, 4, 37, 18, 38].

| Category | Interpretability Method | Output Format |
|---|---|---|
| Inherently interpretable models | Linear regression | Mathematical function |
| | Logistic regression | Mathematical function |
| | Decision trees | If-then rule(s) |
| | Decision rules | If-then rule(s) |
| Surrogate models | Local Interpretable Model-agnostic Explanations (LIME) | Any output format of inherently interpretable models, e.g., mathematical function or if-then rules |
| Feature attribution | Shapley values | Mathematical function |
| | SHapley Additive exPlanations (SHAP) | Mathematical function |
| | Counterfactual explanations | If-then rule(s) |
| | Anchors | If-then rule(s) |

TABLE II. TYPES OF ML PROVENANCE DATA RELEVANT TO INTERPRETABILITY. ADAPTED FROM [4].

| **Data** | **Model** |
|---|---|
| • Datasets [21]<br>• Data type<br>• Data source<br>• Data fidelity (real or synthetic)<br>• Data metadata, as applicable (resolution, size/length, contrast, etc.)<br>• Synthetic data metadata, as applicable (prompt, seed, etc.)<br>• Method of collection, collector, date of collection [21]<br>• Inclusion/exclusion criteria (person, date)<br>• Preprocessing steps (transformations, cleaning, denoising, resizing, cropping, normalization) (person and date for each) [21]<br>• Labeling (annotation type, annotator, date of annotating) [21]<br>• Feature selection, if relevant (person and date) [21]<br>• Training/validation/testing split<br>• Training/validation/testing dataset class proportions | • Pretraining provenance, if applicable and available (model comparison and selection) (person and date)<br>• Model size<br>• Model speed<br>• Parameters (weights and biases) [21]<br>• Hyperparameters (number of layers, learning rate, batch size, dropout) [21]<br>• Architecture comparisons<br>• Optimization algorithms [21]<br>• Pipelines [21]<br>• Execution logs and statistics [21]<br>• Environment configurations [21]<br>• Source code (Jupyter notebooks, library dependencies) [21] |

ML provenance, just like regular provenance, is a key aspect on increasing the authenticity, and trust in an ML model by increasing the chances of reproducing the same conditions the model had when training, and making the results as reproducible as possible [19]. Without being able to get similar results as the creators, there is no way to guarantee that the behavior demonstrated by the model is not an outlier instead of the norm.

### III. APPROACH AND VERIFICATION

This approach uses ML provenance as a lower-level NFR to decompose interpretability, then decomposes provenance into verifiable FRs by defining what provenance data a system will collect. The verification of these FRs then by extension verifies that the provenance and interpretability NFRs have been met for the system. Given that interpretability is the ability for humans to draw connections among model data, training, and behavior, the storage of model provenance throughout development allows for humans to trace behavior from training data to model deployment, making the model interpretable. For the purpose of verifying this approach, this section includes requirement specifications and implementation for one interpretability method from each format category.

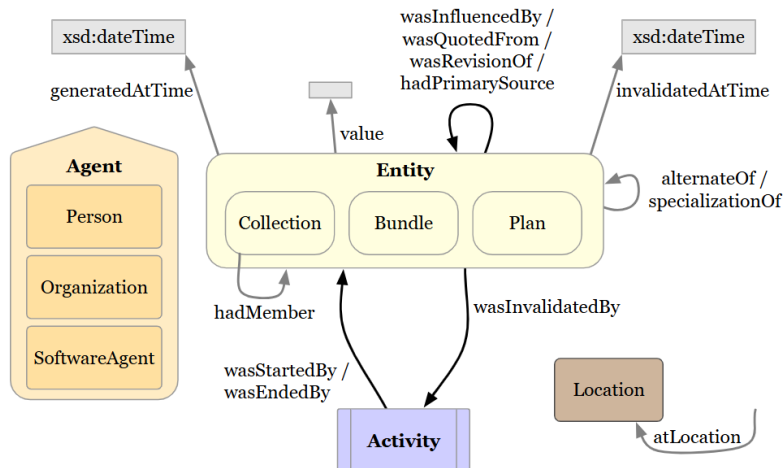

Figure 1. The three "Starting Point" classes and some of their subclasses in PROV-O [28].

```
yprov4ml.start_run(
    prov_user_namespace="www.example.og",
    experiment_name="linear_regression",
    provenance_save_dir="/save/path/"
)
yprov4ml.log_dataset(train_data, "train_data")
yprov4ml.log_dataset(test_data, "test_data")
yprov4ml.end_run(create_graph=True, create_svg=True)
```

Figure 2. Example Python code for saving provenance data.

## A. Provenance Storage

There are several platforms meant to help developers save model provenance including MLFlow, Data Version Control (DVC), Weights & Biases (W&B), Neptune AI, and Comet [22, 23, 24, 25, 26]. Each of these tools can track dataset versions, model versions, hyperparameters, and other common provenance data. However, they are not built to handle more unique provenance data as defined in our previous work, e.g., preprocessing steps or feature selection choices [4]. To accommodate additional necessary provenance tracking, this approach uses yProv4ML, which is compliant with the World Wide Web Consortium's (W3C's) PROVenance Ontology (PROV-O) [27].

PROV-O is an ontology defined using the Web Ontology Language (OWL) that expresses the PROV Data Model (PROV-DM) [28, 29]. It's three "Starting Point" classes, Activity, Entity, and Agent, are shown in Figure 1 along with some of their subclasses [28]. Ontologies represent the structure of entities within a domain [30]. This ontology provides a schema, or structure, for the collected provenance data. yProv4ML then instantiates the structure using PROV-JSON, which can easily be formatted into a graph for readability [27].

yProv4ML is a Python library that handles all of the provenance formatting, making this approach accessible to ML developers, even those without ontology experience [27]. Its usage requires only a small learning curve to know the appropriate function calls for each provenance data type. The output can be viewed either as JSON or as a graph.

## B. Linear Regression Experiment

Linear regression models are inherently interpretable models. Therefore, saving development provenance allows the model's training and predictions to be simulated and verifies the interpretability NFR. This process is shown in this section.

### Requirement Specifications – Linear Regression

The first step is deciding what provenance to save to decompose and verify the interpretability NFR. This provenance may not be the same for all models. The purpose of interpretability is to gather information about the underlying trends in the data and to simulate the model's behavior as a result. To do that for linear regression, one must be able to reproduce the line of best fit for the data and make future predictions using that function. Therefore, the necessary provenance includes the training and testing datasets, the preprocessing steps, e.g., any data trimming, and the resulting function for the line of best fit. This example uses simple linear regression, so these provenance types are sufficient, however additional provenance may be important for multiple linear regression problems, e.g., feature selection. The requirement specifications are as follows:

[Req. 1] The model's development pipeline shall incorporate prediction interpretations for the user.
[Req. 2] The model's development artifacts shall be traceable throughout its lifecycle to ensure interpretability.
[Req. 3] The model's development pipeline shall save development provenance to ensure development artifact traceability to model specifications.
[Req. 4] The model's development pipeline shall save the training dataset to the provenance documentation.
[Req. 5] The model's development pipeline shall save the testing dataset to the provenance documentation.
[Req. 6] The model's development pipeline shall save the data preprocessing steps to the provenance documentation.
[Req. 7] The model's development pipeline shall save the model's linear regression function to the provenance documentation.

### Requirement Specification Verification – Linear Regression

A simple Kaggle example of linear regression is sufficient for the purposes of this experiment [31]. Per the requirement specifications for this experiment, the model's development pipeline needs to save the training and testing datasets, the data preprocessing steps, and the model's linear regression function. Figure 2 is an example of how the training dataset is stored to the provenance documentation.

The full code for this experiment, along with the full provenance JSON file and provenance graph can be found at https://github.com/lynndalou/ProvInterpretExplain.

Two sections of the provenance graph are shown in Figures 3 and 4. Figure 3 shows how the model is saved to the graph and Figure 4 shows the model parameters saved to the graph.

The purpose of saving this provenance data is to interpret and simulate the model's behavior to verify the overarching interpretability NFR. This verification follows the provenance step-by-step to achieve the same result as the Kaggle linear regression example [31]. Microsoft Excel is used for verifying this experimental result. The full Excel file can be found at https://github.com/lynndalou/ProvInterpretExplain. Per the saved provenance data, the following steps were taken:

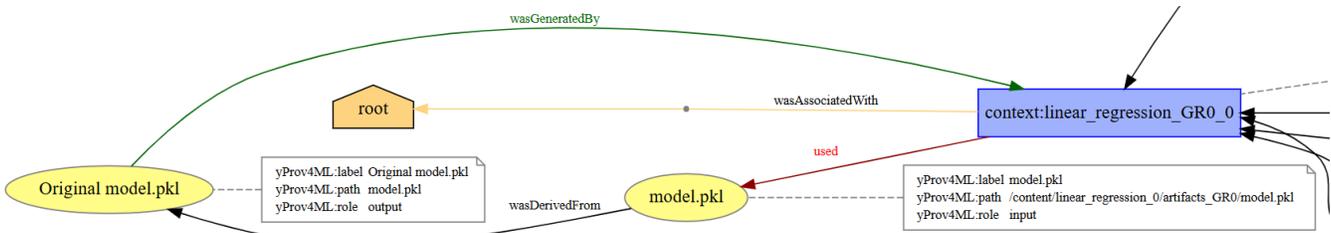

Figure 3. Section of the linear regression provenance graph.

```
dropna_testing_data_cleanup      x    y0  77  79.7751521  21  23.1772792  22  25.6092623  20  17.8573884  ...
dropna_training_data_cleanup     x    y0  24.0  21.5494521  50.0  47.4644632  15.0  17.2186563  38.0  36.58...
execution_end_time               1760554687.6064146
execution_start_time             1760554649.3482456
expand_dims_testing_data         [[ 9.32668894e-01] [-1.00278099e+00] [-9.68219386e-01] [-1.03734260e+00] [-4.84356915e-01] [-1....
expand_dims_training_data        [[-8.99096176e-01] [-4.94443564e-04] [-1.21015062e+00] [-4.15233705e-01] [ 1.27828495e+00] [-4....
standardize_testing_data         [ 9.32668894e-01 -1.00278099e+00 -9.68219386e-01 -1.03734260e+00 -4.84356915e-01 -1.21015062e+00  4...
standardize_training_data        [-8.99096176e-01 -4.94443564e-04 -1.21015062e+00 -4.15233705e-01  1.27828495e+00 -4.84356915e-01 -1...
yProv4ML:artifact_uri            /content/linear_regression_0/artifacts_GR0
yProv4ML:experiment_dir          /content/linear_regression_0
yProv4ML:experiment_name         linear_regression_GR0_0
yProv4ML:global_rank             0
yProv4ML:level                   0
yProv4ML:provenance_path         /content/
yProv4ML:python_version          3.12.12 (main, Oct 10 2025, 08:52:57) [GCC 11.4.0]
yProv4ML:run_id                  0
```

Figure 4. Model parameters saved to the provenance graph.

1. Removing N/A or empty datapoints from the training and testing datasets.
2. Standardizing the independent variable for the training and testing datasets.
3. Using the Analysis Toolpak to calculate the linear regression line [32].

These steps lead to the same linear regression function as the Kaggle code, thereby simulating the behavior of the ML model.

*C. Counterfactual Explanation Experiment*

Inherently interpretable models are unique in that provenance allows their behavior to be simulated, as was shown in the linear regression experiment. Other methods are more complex because they can no longer simulate the model's behavior directly. However, provenance can still be used to map the model's behavior to an input space as a way to understand it. Counterfactual Explanations (CEs) will be used in this experiment to demonstrate how to do this.

Requirement Specifications – Counterfactual Explanations

*[Req. 1] The model's development pipeline shall incorporate prediction interpretations for the user via counterfactual explainations.*
*[Req. 2] The model's development artifacts shall be traceable throughout its lifecycle to ensure interpretability.*
*[Req. 3] The model's development pipeline shall save development provenance to ensure development artifact traceability to model specifications.*
*[Req. 4] The model's development pipeline shall save the training dataset to the provenance documentation.*
*[Req. 5] The model's development pipeline shall save the testing dataset to the provenance documentation.*
*[Req. 6] The model's development pipeline shall save the data preprocessing steps to the provenance documentation.*
*[Req. 7] The model's development pipeline shall save the selected features to the provenance documentation.*
*[Req. 8] The model's development pipeline shall save the counterfactual explanations to the provenance documentation.*

Requirement Specification Verification – Counterfactual Explanations

This experiment utilizes a simple binary classification example with Diverse Counterfactual Explanations (DiCE) [33, 34]. The full code for this experiment and the provenance data saved during the experiment can be found at https://github.com/lynndalou/ProvInterpretExplain. The goal of this experiment is still to understand the black box model's behavior in human-understandable terms, but it is accomplished a bit differently than the linear regression example because the behavior can no longer be simulated. However, provenance data can still help to verify this interpretability requirement.

CEs offer ways to change the inputs to a model that would produce a different output [35, 33]. Therefore, these explanations can be used to map the input space that produces the possible outputs, as is shown in Figure 5. The provenance

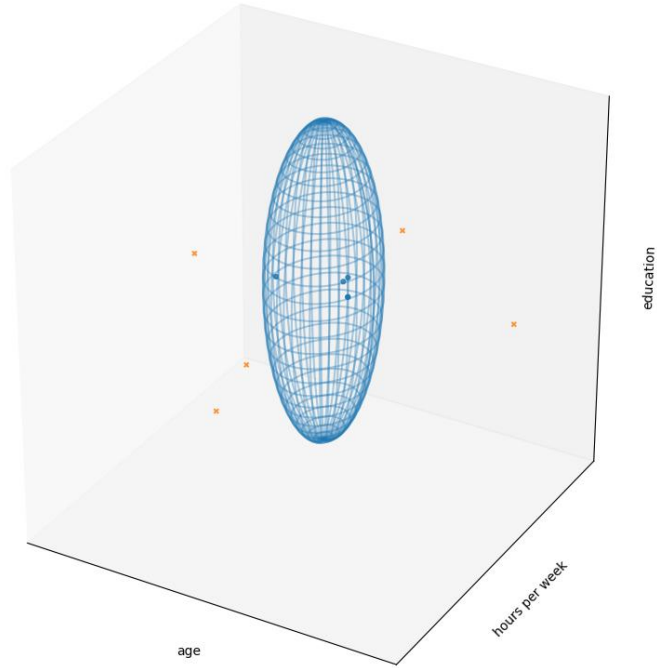
Figure 5. A depiction of how counterfactuals can verify an interpretability requirement.

data required to accomplish this understanding is outlined in the requirement specifications for this experiment. Some of the provenance data is the same as in the first experiment, but this experiment requires feature selection data. As is shown in Figure 5, three of the selected features are "age", "education", and "hours per week". The space then shows "positive" predications, in this case predictions of "high income", within the sphere and "negative" predictions outside of the sphere. Model developers and users can utilize this provenance-based technique to visualize how input changes affect the model's predictions.

## IV. DISCUSSION

These experiments utilize one method from each category in Table I, with the exception of surrogate models because they are also inherently interpretable models, just used in a different way. Each experiment is meant to represent how provenance decomposes an interpretability requirement for any method within their category. Therefore, the decomposition for a decision tree would be similar to that of linear regression because the goal of both is to replicate and simulate the model's behavior. Similarly, the goal of any feature attribution method is to understand how changes in the model's inputs affect the model's predictions, so any method within that category would follow a similar decomposition to CE.

These experiments are proofs of concept, but developers can save additional provenance based on their needs. Figure 5 is also an example visualization that developers and users may find useful in understanding their model's behavior but is by no means the only way to verify interpretability using the saved provenance. Developers may choose to adjust the visualization as needed to fit their needs.

The two experiments work with yProv4ML, which saves most of the data as needed. However, as Figure 4 shows, the preprocessing steps had to be saved as model parameters, which is incorrect. Therefore, this work would benefit from a new provenance tool that could handle more data types, e.g., preprocessing steps.

## V. RELATED WORK

Our previous work has identified provenance as an ML-specific NFR that can decompose interpretability [4]. However, it does not verify the method with experiments, which this work does. Other authors have linked provenance to interpretability or other similar ML quality attributes [5, 36]. Kale, et al. show how provenance contributes to model transparency and by extension, to model explainability [5]. Their literature review discusses how provenance can increase the trustworthiness of ML systems and also mentions how PROV-O and JSON can be used to format and store ML provenance. Although this has greatly influenced this work, it does not show practical examples of how provenance can be used to verify such quality attributes. Meanwhile, Nakagawa, et al. connect provenance and model traceability to ML quality assurance [36]. The authors extend W3C's PROV model to include ML quality characteristics. Although the authors connect provenance to quality verification, this work is distinct in that it frames provenance as a requirement specification that can be used to verify interpretability NFRs.

## VI. LIMITATIONS AND FUTURE WORK

This work successfully verifies the interpretability NFR for two ML models. However, this work does not include a complete survey of interpretability methods. Future work could expand to additional interpretability methods to ensure that this approach verifies all of them.

It is important to note that early difficulties for this work stemmed from confusion in literature as to the exact differences between interpretability and explainability. Future work will compare and contrast the two ML NFRs and analyze how provenance can be used to verify each of them.

Finally, it was noted in the approach section that none of the available ML development and provenance tracking tools were capable of correctly tracking all of the necessary provenance for this work, e.g., data preprocessing steps. Therefore, future work will include the development of a more complete ML provenance tracking tool that is compatible with other development platforms.

## VII. Conclusion

The purpose of this work is to provide a method of measuring and verifying interpretability NFRs for ML-based systems. This is done by decomposing the interpretability NFR into a provenance requirement and then specifying FRs based on what provenance data to save. Developers can then use the saved provenance to simulate or map the black box model's behavior in human-understandable terms.

This method was verified in two experiments. The first used provenance to verify interpretability for a linear regression model. Because this model is inherently interpretable, the goal of saving provenance was to replicate and simulate the model's behavior, which was successfully accomplished. In contrast, the binary classifier in the second experiment could not be replicated and simulated, but provenance was still helpful in mapping the classifier's decisions to the input space to understand how changes in the input features affected the model's predictions. In each case, the provenance data was used to interpret the model's behavior and thereby verify the interpretability NFR. Future work includes verifying that this approach works for additional interpretability methods and expanding this research to encompass how provenance may also verify explainability requirements.